\documentclass{tcibook}
\usepackage{fancyhea}
\usepackage{work}
\usepackage{bm}       
\usepackage{graphicx}
\usepackage{multirow}
\usepackage{lineno}
\usepackage{threeparttable}
\usepackage[normalem]{ulem}
\usepackage{color}
\usepackage[colorlinks = true,
            linkcolor = blue,
            urlcolor  = blue,
            citecolor = blue,
            anchorcolor = blue]{hyperref}

\usepackage{titlesec}
\titleformat{\chapter}{\centering}{}{0em}{\bf\huge}

\def\beq{\begin{equation}}
\def\eeq#1{\label{#1}\end{equation}}
\def\eeqn{\end{equation}}

\newenvironment{Eqnarray}%
   {\arraycolsep 0.14em\begin{eqnarray}}{\end{eqnarray}}
\def\beqa{\begin{Eqnarray}}
\def\eeqa#1{\label{#1}\end{Eqnarray}}
\def\eeqan{\end{Eqnarray}}

\let\bar=\overbar

\def\lsim{\mathrel{\raise.3ex\hbox{$<$\kern-.75em\lower1ex\hbox{$\sim$}}}}
\def\gsim{\mathrel{\raise.3ex\hbox{$>$\kern-.75em\lower1ex\hbox{$\sim$}}}}

\def\W{{\cal W}}

\def\del{\partial}
\def\Dslash{\not{\hbox{\kern-4pt $D$}}}
\def\dslash{\not{\hbox{\kern-2pt $\del$}}}
\def\pslash{\not{\hbox{\kern-2pt $p$}}}
\def\ETmiss{\not{\hbox{\kern-4pt $E$}}_T}

\def\Dlr{\mathrel{\raise1.5ex\hbox{$\leftrightarrow$\kern-1em\lower1.5ex\hbox{$D$}}}}

\def\MSB{{\bar{M \kern -2pt S}}}
\def\msb{{\bar{\scriptsize M \kern -1pt S}}}

\def\drb{{\bar{\scriptsize D \kern -1pt R}}}

\def\authorlist#1#2{
\begin{center}\begin{large} { #1 } \end{large}
    \vskip 0.2in
    
              #2
     \vskip 0.2in
   \end{center}
}

\begin{document}


\pagenumbering{roman}

\parindent=0pt
\parskip=8pt
\setlength{\evensidemargin}{0pt}
\setlength{\oddsidemargin}{0pt}
\setlength{\marginparsep}{0.0in}
\setlength{\marginparwidth}{0.0in}
\marginparpush=0pt


\pagenumbering{arabic}

\renewcommand{\chapname}{chap:intro_}
\renewcommand{\chapterdir}{.}
\renewcommand{\arraystretch}{1.25}
\addtolength{\arraycolsep}{-3pt}

































 


\chapter{Snowmass Topical Group Summary Report: \\ IF04 - Trigger and Data Acquisition Systems}
\label{sec:if04}

\authorlist{D. Acosta (\textit{Rice University}), \\ A. Deiana (\textit{Southern Methodist University}), and \\W. Ketchum (\textit{Fermilab})}
{with contributions from the community}

\section{Executive Summary}
\label{sec:if04_summary}

A trend for future high energy physics experiments is an increase in the data bandwidth produced from the detectors. Datasets of the Petabyte scale have already become the norm, and the requirements of future experiments---greater in size, exposure, and complexity---will further push the limits of data acquisition technologies to data rates of exabytes per seconds. The challenge for these future data-intensive physics facilities lies in the reduction of the flow of data through a combination of sophisticated event selection in the form of high-performance triggers and improved data representation through compression and calculation of high-level quantities. These tasks must be performed with low-latency (\textit{i.e.} in real-time) and often in extreme environments including high radiation, high magnetic fields, and cryogenic temperatures.

Developing the trigger and data acquisition (TDAQ) systems needed by future experiments will rely on innovations in key areas:
\begin{itemize}
\item pursue innovations in the application of Machine Learning (ML) to TDAQ systems, particularly in the co-design of hardware and software to apply ML algorithms to real-time hardware and in other novel uses to improve the operational efficiency and sensitivity to new physics of future experiments;
\item invest in the design of TDAQ system architectures that leverage new technologies, techniques, and partnerships to enable more intelligent aggregation, reduction, and streaming of data from detectors to higher-level trigger systems and offline data processing; and, 
\item develop improved readout technologies that increase data bandwidth and are capable of operating in extreme environments, while fitting the material and power constraints of future experiments.
\end{itemize} 

Critically, innovations in TDAQ rely on the people and processes behind them, and require investments in those people and infrastructure for R\&D. To that end, we call for:
\begin{itemize}
\item increased effort to build and retain domain knowledge for complex TDAQ systems by reliably supporting facilities and people – particularly engineers and technical staff, and early-career scientists through recruitment and training – in order to bring new ideas from early design and prototyping all the way through integration, commissioning, and operation in future detectors; and,
\item the creation of a dedicated (distributed) R\&D facility that can be used to emulate detectors and TDAQ systems, offer opportunities for integration testing (including low- and high-level triggering, data readout, data aggregation and reduction, networking, and storage), and develop and maintain an accessible knowledge-base that crosses experiment/project boundaries.
\end{itemize}

\section{A compelling use case: track-based triggers}
\label{sec:if04_track_triggers}

Searches for novel new physics are often made possible by the development of novel new triggers that take advantage of improvements in detectors and real-time computing. Early-stage trigger systems of experiments at hadron collider experiments, for example those at the Large Hadron Collider (LHC), often select objects with high momentum, based on calorimeter information, in order to reduce the large backgrounds to events of physical interest.  However, this process necessarily loses many events, and is poorly optimized to explore a variety of potential beyond Standard Model (SM) scenarios.  For instance, low momentum events with displaced vertices could lead to soft-unclustered-energy-patterns, long-lived staus or the decays of long-lived dark scalars in the Higgs portal scenario~\cite{wp_exotracktrig}.  Alternatively, high-momentum and short-lived particles, such as dark matter particulates, might be missed by the existing trigger systems due to their invisible decay within the detector volume~\cite{wp_tracktrig}.

For both of these cases, early-stage triggers based on tracking information could preserve interesting signal events for study.  A difficulty of including tracking information in early-stage triggers is the complexity of calculating tracks and the speed at which decisions need to be made.  For instance, at the LHC~\cite{LHC} events are produced at a rate of 40 MHz, and this must be very rapidly reduced to the kHz level at the first step of the trigger system. The development of a fast track-based trigger is therefore an area of considerable interest in the high energy physics community, and several different approaches have been taken to solve the problem for silicon-strip tracking systems.  
For future hadron collider experiments, a fast tracking trigger for silicon pixel tracking layers at small radii with respect to the beam pipe, and therefore with significantly more channels than strip trackers, would open the sensitivity to particles arising from beyond SM physics in an interesting lifetime regime.

As a new approach, a white paper contributed to Snowmass~\cite{wp_tracktrig} utilizes highly-parallelized graph computing architecture using field-programmable gate arrays (FPGAs) to quickly performing tracking in small-radius silicon detectors.  For LHC experiments such as ATLAS~\cite{ATLAS} and CMS~\cite{CMS}, the silicon tracking detectors consist of multiple cylindrical layers surrounding the proton-proton interaction point.  As particles pass through the detector, they interact with each layer and leave a point of information on their location.  At the HL-LHC, we are expecting up to 200 proton-proton collisions per 25 ns, which is expected to result in approximately 11,000 charged tracks, which in a ten-layer detector would then result in a point-cloud of approximately 110,000 points.  These points would need to be classified, such that each set corresponds to a single charged particle. CPU methods to calculate these tracks are prohibitively slow for use in the trigger, and strategies utilizing associated memories with large pattern banks for matching have met with some success but have not been adopted by LHC experiments.

\begin{figure}[htp]
    \centering
       \includegraphics[width=0.9\textwidth]{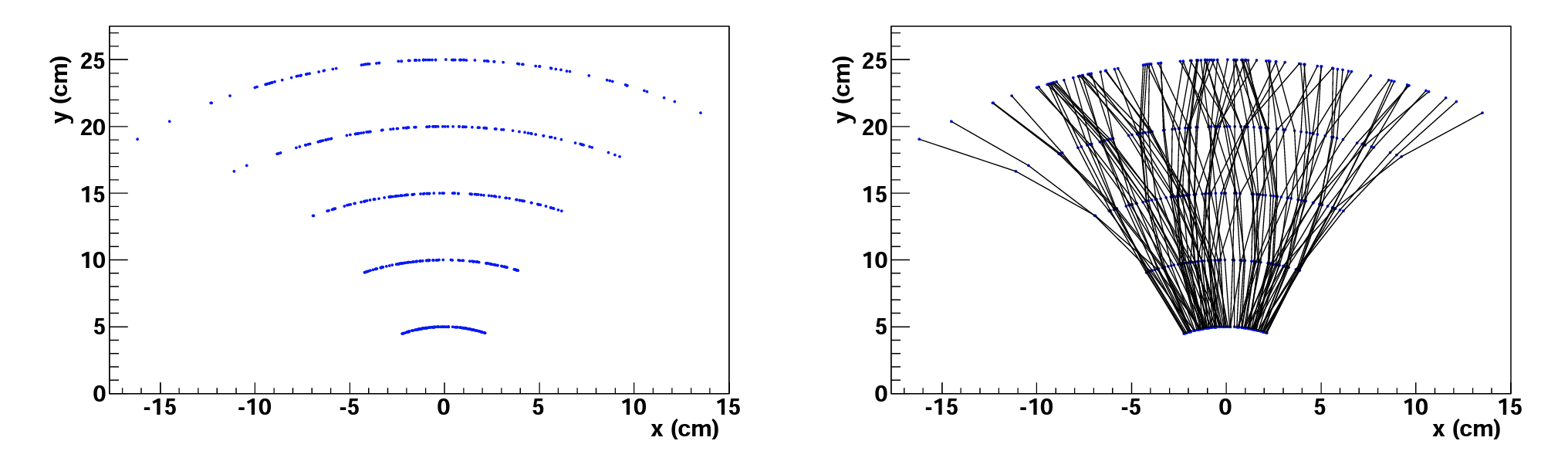}\\
    \caption{(left) An example of the point cloud generated by 100 particles in an azimuthal sector of the silicon pixel detector of width one radian. The silicon sensors are placed in concentric circles with radial separation of 5cm.(right)The reconstruction of 100 tracks from the point cloud.~\cite{wp_tracktrig}}
    \label{fig:tracktrig}
\end{figure}

Instead, this paper proposes unsupervised machine learning on a highly-parallelized graph computer constructed using modern FPGA technology. The method converts a 2-dimensional matrix of points into a graph by defining a weighted link that associates each point with all other points. Then, the tracking problem is reduced to pruning spurious links until the points are divided into groups corresponding to each charged track. The algorithm is tested on a simulated silicon detector with five layers and 100 tracks, with the assumption that each particle is successfully detected on every layer and that there are no noise hits or resolution effects. In this case, track-finding is over 99.95\% efficient, and the layers and reconstructed tracks are illustrated in Fig.~\ref{fig:tracktrig}. This shows that the algorithm is very promising.
%
%
The researchers have taken the next step of shifting to an `imperfect' (i.e. realistic) detector model that includes radius-dependent hit occupancies and noise consistent with HL-LHC simulations. To fit within latency requirements and FPGA resources, the algorithm has been modified by increasing the minimum track transverse momentum. This updated algorithm, documented in \cite{GraphTracking}, yields a trigger efficiency near 100\% for track transverse momenta above 10~GeV and a spurious trigger rate of a few kHz. Implementation studies on a Virtex Ultrascale+ FPGA are in progress and indicate a latency of 200~ns, well below the 4 $\mu$s limit for trigger algorithms at LHC experiments. The algorithm can be tuned for smaller  transverse momenta or different radial granularities/configurations, in order to match longer latency and lower throughput requirements.
Utilizing tracking information at early-stage triggers would be greatly beneficial for a wide variety of physics analyses at LHC experiments and future lepton and hadron collider experiments, so this is expected to remain an area of considerable interest for the future.

The development of future track-based triggers represents a specific example of many of the general thrusts of innovations in TDAQ for future high energy physics experiments. Use of machine-learning techniques to develop artificial intelligence algorithms that are sensitive to new physics, and performing real-time inference through their integration into detector front-ends or FPGAs (for low-level triggers) and/or heterogeneous computing systems using CPUs, GPUs, and other devices (at high-level) is a rapidly improving and highly promising direction for a wide variety of TDAQ applications. Demonstrating these types of triggers in current and near-future collider experiments (at the LHC and HL-LHC) may open the door for exploring new physics at more distant-future colliders, like the FCC, particularly when considered alongside further innovations in improved readout technologies that can increase data bandwidth and data locality.

Finally, we note that the design of future detector systems, especially tracking systems in dense, high occupancy environments, is best done taking into account the necessary trigger and DAQ considerations early in the process. For example, the design of the HL-LHC silicon tracker for the CMS experiment \cite{CMSPhase2Tracker} utilizes the local coincidence of hits in two closely-spaced tracking layers to provide some limited momentum selection capability. This is necessary to  reduce the data rate from the detector to manageable levels for the early-level trigger system.

\section{Heterogeneous Computing Hardware and Machine Learning}
\label{sec:if04_heterogeneous_and_ML}

As is well known, the scaling of single CPU solutions to efficiently address computational problems has reached its limits, and the need for the parallelization of tasks across many CPU cores, in Graphical Processing Units (GPUs), or in other specialized hardware like Field Programmable Gate Arrays (FPGAs) and Application Specific Integrated Circuits (ASICs) has become necessary. (The choice of technology often depends on the timescale and amount of data to handle). For high-energy physics experiments this means moving beyond the natural parallelization of processing individual data ``events'' (e.g. beam crossings, which are independent of each other) to the parallelization of the algorithms themselves acting upon the data of a single event. This requires a new paradigm of coding algorithms to take advantage of the available heterogeneous computing hardware, and thus utilizing industry tools or developing domain specific tools to aid in this parallelization. Furthermore, evolution of computing hardware may involve tradeoffs between overall compute power, latency, and other resources. There is a need for continued evaluation of next generation processor technologies to understand these tradeoffs, and how that may impact TDAQ system design.

Machine learning algorithms lend themselves well to be distributed on such heterogeneous computing platforms using standard libraries, and thus make a natural and powerful target for trigger applications. The applications include the very specialized and local processing at the front-end of the detector electronics (``edge'' computing) where low-level detector hits are converted into clusters or other higher-level data objects at high frequency and low latency, but also the more global and generalized processing needed to discriminate physics signatures from backgrounds. The latter might include discriminating low-energy signal events from backgrounds in neutrino experiments, Higgs boson decays in collider experiments, or gravitational wave signatures for multi-messenger astrophysics. Machine learning also could potentially be used to go beyond the fixed hand-curated trigger menus used to select physics data at colliders to a novel ``self-driving'' paradigm whereby the trigger system autonomously and continuously learns from the data to more efficiently and effectively filters and selects data from a detector system, as discussed in white paper \cite{TDAQ-innovations}. Such systems may complement dedicated triggers, searching for specific signatures of Beyond-Standard-Model physics, by performing a general set of anomaly detection that may be sensitive to a wider variety of new physics.

\subsection{Fast Machine Learning in early stage trigger}
\label{sec:if04_FastML}

Development of machine learning algorithms for use in the trigger and data acquisition systems of future high energy physics experiments is particularly challenging~\cite{fastml_summary}. For example, while ML-based algorithms have proven effective at performing effective data reduction, through both advanced data selection and data compression, to be used in high-luminosity environments of future particle colliders these algorithms must be capable of running in on-detector electronics with latencies on the order of nanoseconds. Future neutrino, dark matter, and astrophysics experiments require latencies on the order of microseconds to milliseconds, and often similarly extreme detector environments.

The design and development of low-latency AI algorithms requires optimization across both physics (\textit{e.g.} selection efficiency, background rejection) and technical performance (\textit{e.g.} latency, resource usage). It's necessary to consider a co-design of hardware and software that gives special attention to the processor element, making use of tools and expertise that can bring a variety of ML algorithms built on large datasets into FPGA and ASIC firmware. Open-source frameworks like \texttt{hls4ml}~\cite{hls4ml} and FINN~\cite{FINN} aim to ease the complexity of firmware programming, which have opened up development and integration of sophisticated AI into high-performance hardware. Continued development of ML frameworks that can aid hardware/software co-design, coupled with (and in many cases driving) improvements in the underlying processor technologies, can open the door to paradigm shift in how future experiments will collect, reduce, and process data.

\subsection{Fast Machine Learning development strategies}
\label{sec:if04_FastML_dev}

To facilitate fast ML development, it's also very important to foster interdisciplinary collaboration between electrical engineering, computer science and physics, as people in these fields have valuable expertise in digital design, machine learning techniques and the physical problems to be addressed. The community needs to keep track of and be willing to adapt to the latest developments in industry, but also retain the expertise to continue developments addressing challenges that go beyond industry standard. An important note is the value of keeping active work, as much as possible, open source, though the main distributors of FPGAs in the current market (Xilinx and Altera) require the use of proprietary development software. As much as possible, work should be preserved in an open source manner, to be used cross-project and cross-experiment and further built upon in the future.



\section{Innovative Architectures}
\label{sec:if04_architectures}

\subsection{Time Multiplexed or Asynchronous Hardware Trigger Systems}
\label{sec:if04_async}

For experiments where a processing element of the trigger system must have a complete view of the data from all detectors to perform its function, scalability of the trigger system can be achieved by time multiplexing the data to individual processors. This is natural for software-based trigger levels (CPUs, GPUs) in collider experiments, where data are aggregated in an event builder and sent to a target compute node for processing asynchronously. For a hardware-based synchronous trigger level (FPGAs, ASICs), this can be achieved by sending data from all detector elements for a given time slice (or event)  via links to target processors in a round-robin fashion (``time multiplexing''). This will be necessary for the CRES tritium beta decay experiment, for example, where each compute node must process the data from all receivers in a given time slice.

In collider experiments, trigger data processing in the hardware-based level is still generally synchronous to the accelerator clock, even if the processing is time multiplexed. At all stages the data are processed and registered at a multiple of this frequency in the digital pipeline, and the event number is implicit by the clock (or accelerator bunch) counter. However, it is enough to time-stamp data at the very front end of the detectors with the system clock, and transmit and process the data asynchronously as traditionally done at the software-based level~\cite{async-L1}. Effectively the event builder infrastructure moves to the first level of the trigger system. This would alleviate the challenge of distributing and synchronizing a stable, low-jitter, high frequency clock  over the entirety of a very large and distributed electronics system. It only needs to go to where the data are marked, not to all of the processing elements of a trigger system. It also has the additional benefit of blurring the lines between the first level trigger, which typically runs in fast FPGAs, with the software-based higher levels. It might be possible in the same compute node to run first the very fast algorithms on one set of attached (FPGA) resources,  with subsequent processing on other attached GPU or CPU resources on the same node for more complex software-based algorithms. Another avenue of potential interest in this area is to make use of neuromorphic computing directly on analog signals. 

\subsection{Streaming DAQ}
\label{sec:if04_streaming_daq}

Another innovative approach to solving the data reduction problem for trigger and data acquisition systems is to move away from a pipelined and triggered readout, and instead operate in a more ``streaming" design, where data is encoded with its time and origin~\cite{TDAQ-innovations}. In this model, pioneered by the LHCb experiment for its upgrade but being adopted to some degree as well by the other LHC experiments, event data can be reduced at its source, often through simple thresholding and zero suppression, and then aggregated and streamed to downstream computational and storage elements. There, full- or partial-event filtering and further processing and translation of data into higher-level quantities can be performed in order to achieve the reduction in the data throughput and offline computing. Hybrid designs that combine both traditional trigger-based DAQ for some detector subsystems and streaming-readout for others is also possible. Emphasis on such approaches for upgrades and experiments at future facilities has merit, especially due to its ability to simplify DAQ design.

\section{Developments in novel readout technologies}
\label{sec:if04_readout}

The needs of future detectors continue to push readout, triggering, and data acquisition technologies to operate with growing data rates in more extreme environments. Future kton-scale neutrino and dark matter experiments like DUNE and LZ will produce many petabytes of data per year with intrinsic data rates in excess of TB/s, and require readout systems that can reliably operate in cryogenic temperatures over the long lifetimes of the experiments and minimize radiological material volumes to maintain sensitivity to low-energy interactions. In high-energy collider physics at the HL-LHC or potential future colliders like the FCC-hh, data rates in the hundreds of TB/s are possible from tracking and calorimetry systems, and must be able to withstand high radiation rates and not significantly add to the material budget of the detectors. The fundamental challenge of how to move data from readout electronics to online and offline computing resources requires a commitment to research and development in new and improved technologies.

Core to improvements in DAQ are a combination of reducing the data rate close to the detector, and increasing the data bandwidth for a given material and/or power cost, in the extreme environments required. More sophisticated data reduction techniques in detector electronics may be possible with advances in AI, particularly with improvements in translating low-latency machine-learning-developed compression and triggering algorithms to ASICs~\cite{fastml_summary}. In many cases, like in fast tracking algorithms, correlations across different portions of the detector are necessary to develop effective trigger algorithms, and thus fast, localized, and low-material communication is necessary. Wireless communication technologies are an area of large promise here: microwave-based technologies already show reliable data transmission at the 5 Gb/s scale, and promising future work using free space optics may allow for wireless communication at the Tb/s scale \cite{readout-innovations}. Integration of wireless communication into HEP detector design (like, for future tracking detectors in colliders), could allow for new system designs that exploit localized readout, fast analysis, and triggering to intelligently reduce data volumes.

Technologies to allow greater bandwidth off of the detector also show significant promise~\cite{readout-innovations,readout-links}. Silicon-photonics are an appealing alternative to the current VCSEL-based approaches: they can allow integration of fiber-optic connections directly to sensor modules or readout chips (thus reducing the need for electrical cable connections), commercial devices already show high radiation tolerance, and offer a bandwidth twice that of VCSEL devices with a power consumption that is 20\% less. More promising developments exist in Wavelength Division Multiplexing (WDM), where individual serial links can be transmitted on its own wavelength, reducing the need for data aggregation to maintain high data bandwidth per link. WDM could be used in a design to bring data out of the most extreme radiation environments in colliders more efficiently, allowing for further data reduction in downstream DAQ components.

Notably, the work to develop new readout technologies is often less in design, but more in integration and testing with real detector components in real detector environments. It is important to develop and maintain tools and facilities that can allow for realistic full-system testing of readout electronics and DAQ. 

\section{Timing}
\label{sec:if04_timing}

Time measurements will feature prominently in the next generation of particle physics experiments and upgrades, including integration into ``4D'' tracking systems 
and ``5D'' calorimeters
. Timing helps disentangle the effects of particle pile-up in hadron collisions, discriminates against beam-induced backgrounds in a muon collider experiment, and can be used to separate Cerenkov and scintillation light signals in neutrino and dark matter experiments. Timing also provides particle ID information useful for a broad range of experiments, including sensitivity to any slow beyond-standard-model long-lived particles. Thus it seems evident that timing information will work its way into the trigger processing chain to improve its selectivity and precision of measurements. A particular challenge is the need for synchronization of data at the ${\cal O}(10)$~ps level or better, including across large distances in the case of radio-frequency arrays.


\section{Support for installation, commissioning, integration, and operations}
\label{sec:if04_support}

Success of a detector upgrade or new experiment ultimately depends not just on the delivery of the new components, but on the successful installation, commissioning, and integration of them into the experiment as well as their efficient operation. This applies to trigger-DAQ as much as any other instrumentation area~\cite{TDAQ-innovations, readout-innovations}. These tasks need to be well thought out (preferably as part of the initial proposal to evaluate the overall cost of a new system) and supported. Lack of attention in any of these areas can lead to substantial and costly delays as well as a failure to reach the design goals, which jeopardizes the physics output. Thus the installation, commissioning, and integration tasks as well as the long-term operations (until the end of the experiment) must be a priority. 

\section{Building and retaining expertise in TDAQ instrumentation}
\label{sec:if04_community}

Along with the importance in conducting the R\&D to develop and ultimately construct innovative trigger and data acquisition systems for future physics facilities, equally important is to build and retain the domain knowledge and technical expertise within the high-energy physics community required to support this~\cite{TDAQ-innovations}. It can be challenging to recruit and retain highly skilled personnel to address the specialized and high-tech needs of our community. Essential technical staff can leave for higher-paying positions in industry, and younger scientists specializing in instrumentation may/will find career progression and promotion a challenge in this field. This is compounded by the timescale for large experiment facilities from construction through to the end of operations, which can be decades. Thus it is imperative for the scientific community to provide career opportunities for such highly skilled people. 

Retention of highly-skilled personnel is made even more important because the needs of particle physics experiments are not always relevant to industrial partners.  The particle physics community has an interest in edge cases to typical uses in industry, such as electronics that perform well under constant bombardment from radiation or in cryogenic fluids. There's also a need for a high degree of reliability, as many devices are installed in detectors in areas that are inaccessible for replacement for decades at a time.  It's important for the community to continue to follow development in industry and to build strong collaborative networks, but it is as important to develop resources to pursue R\&D directions that are specific to use cases in the particle physics domain.

As we explore new TDAQ architectures and hardware with increasing complexity whose performance depends on interactions on a systems level, a dedicated facility that can support TDAQ design and development while also offering opportunities for integration testing across low- and high-level triggering, data readout, data aggregation and reduction, networking, and storage should be established. While the necessary support hardware for the facility may be localized, the participating domain experts should encompass a distributed community. Given many of the common challenges across physics frontiers, such a Trigger and Data Acquisition Emulation and Integration Test Facility should cross experiment and project boundaries, offering support for emulation of detectors and TDAQ systems, and the development and maintenance of common hardware, firmware, and software, and  to support TDAQ R\&D for future detectors. This facility will develop and maintain an accessible knowledge-base by enabling and supporting connections and communication between engineers and scientists in many national and university labs working within different sub-fields that encounter similar problems.  









\end{document}